\documentstyle[12pt]{article}
 \def\L{{\cal L}}  \def\1{{\bold 1}}
    
\def\d{\partial}    
\def\alfa{\alpha}    \def\R{{\bf R}}  
\def\eqdef{\buildrel {\rm def} \over =}     
\def\({\left(}    \def\){\right)}
      
\def\colon{\quad;\quad}    
   
\def\data{\the\day-\the\month-\the\year}

  \def\phidot{{\dot \phi}} 
    
\def\D#1{\frac{\d}{\d #1}} \def\frac#1#2{{#1 \over #2}}

\def\phi{\varphi}

\def\half{\frac{1}{2}} 
 \def\ff{\frac{3}{2}} 

\def\beqa{\begin{eqnarray}}
\def\eeqa{\end{eqnarray}}
\def\num#1{(\ref{#1})}
\def\beq{\begin{equation}} 
\def\eeq{\end{equation}}  \def\bib#1{[\ref{#1}]}

 \def\n{N_1} \def\nn{N_2} \def\nnn{N_3} \def\b{\beta_1}
 \def\bb{\beta_2} \def\l{\lambda}  \def\RR{R^*}
 \def\ld{{\dot \lambda}} \def\bd{{\dot \beta_1}}  
 \def\bbd{{\dot\beta_2}} \def\om{\omega}
\def\jmp{{\it J. Math. Phys.}\ }
\def\pr{{\it Phys. Rev.}\ }

\def\pl{{\it Phys. Lett.}\ }

\def\ijmp{{\it Int. Journ. Mod. Phys.}\ }

\def\cqg{{\it Class. Quantum Grav.}\ }

\def\prs{{\it Proc. R. Soc.}\ }

\def\apj{{\it Ap. J.}\ }
\def\aa{{\it Astron. Astrophys.}\ }
\def\ncim{{\it Nuovo Cim.}\ }

\def\prep{{\it Phys. Rep.}\ }

\begin{document}
\begin{titlepage}
\title{\bf  N\"other symmetries in Bianchi universes }
\author{{S. Capozziello, G. Marmo, C. Rubano, P. Scudellaro} %
\\{\em Dipartimento di Scienze Fisiche, Univ. di Napoli and INFN, Sez.
di Napoli}
\\{\em Mostra d'Oltremare, pad. 19, 80125 Napoli.
Italy } }
 \maketitle
\begin{abstract}
        We use our N\"other Symmetry Approach to study the Einstein
equations minimally coupled with a scalar field, in the case of
Bianchi universes of class A and B. Possible cases, when
such symmetries exist,  are found and two examples of exact integration 
of the equations of motion are given in the cases of Bianchi AI and BV. 
 \end{abstract} \vfill
 capozziello@na.infn.it; 
gimarmo@na.infn.it;
rubano@na.infn.it;
scud@na.infn.it
\end{titlepage}

 \section{\normalsize \bf Introduction}
Astronomical observations imply that, on a very large
scale, our universe can be simply considered as homogeneous and
isotropic for cosmological purposes, being thus well described by
a Friedman--Robertson--Walker (FRW) metric \bib{kolb}. It
was suggested \bib{hoyle} that such a scenario could be
nothing else but the result of a different previous history and,
therefore, be also consistent with a large amount of primordial
anisotropy.

If, moreover, one adopts an inflationary scenario to describe
the very early history of the universe, it can be expected that
what is actually observed today does not depend on specific
initial conditions, including possible initial anisotropies \bib{wald}. 
On the
contrary, pre--inflationary anisotropies can play a role if primordial 
friction
coefficients, responsible for the coupling between the gravitational
field and the inflaton field, are supposed. As a matter of fact, such 
 coefficients are proportional to the rate of expansion of
volume, and their increasing makes inflation work better (see for
example \bib{rothman}).

Furthermore, since we generally expect that all homogeneous
anisotropies do not increase, it is not strange that we  find
very small anisotropies in the microwave background 
$(10^{-7} \div 10^{-4})$ on large and small angular scales
\bib{kolb}, \bib{barrow}, \bib{smoot}.  Anyway, it is possible to conjecture
that anisotropic cosmological models isotropize as time approaches
our epoch \bib{barrow}, \bib{collins}, which makes it meaningful to study
situations in which isotropy is not necessary at early times.

On the other side, as we just mentioned before,
introduction of a scalar field in cosmology is widely used, 
to obtain  inflationary eras at the very beginning and to treat
primordial matter fields in the schemes of Grand Unified Theories.

For all these reasons, it is worthwhile to understand
dynamics of anisotropy and to search for anisotropic cosmological models
which are exactly solvable.

In this paper, we  investigate 
homogeneous theories of gravity, in which a scalar
field is minimally coupled to gravity, 
searching for point symmetries in the cosmological Lagrangian
density which allow to solve exactly  the dynamical problem
(we begun such a program in \bib{demianski}).

We start from the action
\beq
\label{1}
{\cal A} = \int{d^4 x \sqrt{-g} \left( R + \half g^{\mu \nu} \phi_{;\mu}
        \phi^{;\nu} - V(\phi) \right)} \ ,
\eeq
where $R$ is the Ricci scalar curvature for a spatially
homogeneous spacetime, described by the metric
\beq
ds^2 = - dt^2 + g_{ab}(t) \om^a \om^b \,,
\label{2}
\eeq
$\om^a $ being $1$-forms, not necessarily closed ($
a,b = 1,2,3$). By $\phi_{;\mu} $ ($\mu = 0,1,2,3$) we denote the
covariant derivative of the scalar field  $\phi$; $g$ is the
determinant of the metric, and $V(\phi)$ is the potential 
for $\phi$. 

Brans-Dicke gravitational theories have been recently \bib{Jantzen} 
examined in a wider context. Anyway, a search for point symmetries in 
vacuum situations, in which 
 $\phi$ is  nonminimally coupled to gravity as well,
has been recently performed in \bib{lidsey} in a form, however, which is
not immediately reducible to the minimally coupled ones.

Here we search for N\"other symmetries of  Lagrangian \num{1},
 in  absence of ordinary matter, in order to select all
Bianchi models which are exactly solvable by the so called
{\it N\"other Symmetry Approach}
\bib{demianski}, \bib{deritis}. 

The paper is organized as follows.
The Bianchi classification and the
Lagrangian description are sketched in Sect.2. 
 In Sect.3 we apply our procedure to
determine the cases where symmetries are present. Sect.4 is
devoted to the analysis of the results, and in Sect.5 we give two
examples of exact integration of the equations of motion. 
Conclusions are drawn in Sect.6. 

\section{\normalsize \bf Bianchi classification and Lagrangian formulation}
Bianchi spacetimes, in modern geometric language, are $G_{3}-$principal
fiber bundles over $R$. The bundle being trivializable is usually written as
$M^4 = R \times G_3 $. $G_3$  stays for the three--dimensional
group of
isometries of the metric on $M^{4}$.
If one is not paying attention to the topology of $G_{3}$
(i.e. not distinguishing between the simply connected group with
respect to the quotient groups obtained by quotienting it with discrete
normal subgroups (not distinguishing among Lie groups with the same
Lie algebra) we can use a parallelization of $M^{4}$ with the vector fields
$X_{0},X_{1},X_{2},X_{3}$ with $X_{0}$ the "time"--generator and
$X_{1},X_{2},X_{3}$ the Killing vector fields of the metric closing on the
Lie algebra of $G_{3}$. Of course, when $X_{1},X_{2},X_{3}$  are
known (and complete) we can integrate them to the action of $G_{3}$
on $M^{4}$.

        In the framework of this procedure, in 1897 Bianchi
worked out the first classification of three--dimensional Lie
algebras \bib{bianchi}. Of course, metrics for which corresponding groups
are isometries describe cosmological models with specified symmetry
properties. Bianchi started with the adjoint 
representation and his approach has been followed since then. 
Here, we
would like to comment briefly on the classification by using the 
coadjoint representation, as recently proposed in \bib{marmo1} and 
\bib{marmo2}. 

        As it is well known a Lie group is locally characterized by its 
structure constants  $c^a_{bc} $. If $(X_a)$ is a basis for the vector
space underlying the algebra ${\cal G}$ we have $ [X_a,X_b] = c^c_{ab}X_{c} $ 
and 
the Jacobi identity $ [X_a,[X_b,X_c]] + [X_b,[X_c,X_a]] + [X_c,[X_a,X_b]] 
= 0 $ gives a quadratic constraint on the structure constants. 

By identifying the vector space ${\cal G}$ with the linear 
functions on ${\cal G}^*$, $ {\cal G} = {\rm Lin}({\cal G}^*,\R)$, 
we can define a Poisson 
bracket on ${\cal G}^*$ by setting $ \{ x_a,x_b\}  = c^c_{ab} x_c $. The 
classification problem is reduced now to the problem of classifying 
linear Poisson brackets on a three-dimensional vector space. By using 
the transformation properties of a Poisson bracket under change of 
coordinates one shows that with any bracket is associated a bivector 
field $\Lambda$. In our case $ \Lambda = c^c_{ab} x_c \D{x_a}\wedge\D{x_b} 
$. 

In three dimensions, any bivector field is associated with a 
one-form $\alpha$ by means of the contraction with a volume form $ 
\Omega$, say $ \Omega = dx_1 \wedge dx_2 \wedge dx_3 $. Thus, we have 
to consider and classify quadratic one-forms $ \alpha = c^c_{ab}x_c 
\epsilon^{abr} dx_r = H^{cr} x_c dx_r $. The Jacobi identity is 
equivalent to $\alpha\wedge d\alpha = 0$ \bib{marmo1}. 

The one-form $\alpha$
can be written by decomposing $ H^{cr}$ into symmetric and 
skew-symmetric part 
\beq
\alpha = \frac{1}{2} H^{(cr)} (x_r dx_c + x_c dx_r) +
         \frac{1}{2} H^{[cr]} (x_r dx_c - x_c dx_r) \ .
\eeq
By using linear transformations on the three dimensional vector space,
we can bring $\alpha$ to the normal form
\beq
\alpha = \frac{1}{2}d(\n x_1^2 + \nn x_2^2 + \nnn x_3^2) +
\frac{a}{2}(x_2 dx_3-x_{3}dx_{2}) \ .
\eeq
Imposing the Jacobi identity, we find $ a N_1 = 0 $
and, by means of a suitable change of coordinate basis, we may set
 $ N_i = 0, -1, +1 $.
Thus all 
quadratic one-forms giving rise to three dimensional Lie algebras are 
classified in two classes
\begin{enumerate}
\item Bianchi A , $\alpha$ exact, i.e. $ a = 0 $
\item Bianchi B, $ d \alpha \neq 0 $, i.e. $a\neq 0$.
\end{enumerate}

In this approach, class A is associated with a foliation of $\R^3$ 
given by the Casimir function $ \n x_1^2 + \nn x_2^2 + \nnn x_3^2 $. 
The level sets of this function are the coadjoint orbits of the Lie 
group $G_3$ associated with the Lie algebra ${\cal G}$. They give information
on the topology of the group. Clearly, if $ \n \nn \nnn \neq 0 $, we 
have either $O(3)$ or $O(2,1) $, according to the signs of $N_i N_j$
(the second, if one of them is negative). If at least one of the $N$ is
zero, the group is noncompact.

 The parameters $a, N_i$ are sufficient to
characterize Bianchi classification, according to the following
table

\begin{center}
{\bf Table I}--Classification of Bianchi types\\
\vspace{1. mm}
\begin{tabular}{||l|l|l|l|lr||} \hline
{\it Type} & $a$ & $N_{1}$ & $N_{2}$  & $N_{3}$ &\\ \hline
      I    & 0   &  0      &     0    & 0        &\\ \hline
      II   & 0   &  1      &     0    & 0        &\\ \hline
      III  & 1   &  0      &     1    & -1       &\\ \hline
      IV   & 1   &  0      &     0    & 1        &\\ \hline
      V    & 1   &  0      &     0    & 0        &\\ \hline
 VI$_{o}$  & 0   &  1      &    -1    & 0        &\\ \hline
 VI$_{h}$  & $a$ &  0      &     1    & -1       &\\ \hline
 VII$_{o}$ & 0   &  1      &     1    & 0        &\\ \hline
 VII$_{h}$ & $a$ &  0      &     1    & 1        &\\ \hline
    VIII   & 0   &  1      &     1    & -1       &\\ \hline
    IX     & 0   &  1      &     1    & 1        &\\ \hline
\end{tabular}
\end{center}

A parameter $h \eqdef a^2/(\nn\nnn) $, is also introduced for
subclassifying tipes VI and VII. 

Because any Lie group is parallelizable, we can use left-invariant (or
right invariant) one-forms in the metric (\ref{2}), so that 
it will be invariant under left (or right) action of $G_3$ on
$M^4$. By using a diagonal form for $ g_{ab} $ 
and selecting (for obvious reasons)
a positive definite one, we can parametrize $g_{ab}$ by factoring 
$\det \Vert g_{ab} \Vert $, namely, following Misner, 
we obtain \bib{ellis}, \bib{mccallum}, \bib{ryan}
 \beq g_{ab}(t) = e^{2
\lambda(t)} e^{2 \beta_{ab}(t)} \ , \label{6}
 \eeq
where the matrix $\beta$ is diagonal and traceless, thus depending
on two variables only. A widely used choice is \bib{ellis}
 \beq
\Vert \beta_{ab} \Vert = {\rm diag}\left( -\half \b +
\frac{\sqrt{3}}{2} \bb \ ,\  -\half \b - \frac{\sqrt{3}}{2} \bb \ ,\
  \b \right) \ .
  \label{7}
  \eeq
  
Being $ \sqrt{-g} = e^{3\l}$, the expansion of volume is wholly
determined by $\l$. As to the shear, it is determined only by the
$ \beta_i $. This is typical only for the `orthogonal' case, when
there are no effects of rotation and tilt. Spatially homogeneous
sections are then orthogonal to the fluid flow vector in the
universe \bib{ellis} \bib{mccallum}. 

As shown in  \bib{ellis}, the second--order Einstein equations of all
Bianchi types, except IV and VII$_{h \neq 0}$, can be derived by a
point Lagrangian. In Sect. 5, we shall  consider  also  the first
order Einstein equation, which is always a first integral of the
second order ones. 

Then the general expression for the Lagrangian is the following 
\beq \L = e^{3\l} \left( \RR + 6 {\dot \l}^2 - \frac{3}{2}
( \bd^2 + \bbd^2 )
 - \phidot^2 + 2V(\phi) \right) \ ,
\label{8} 
\eeq
 with the scalar curvature of the spacelike hypersurfaces given by
\beqa
 \RR &=&-\half e^{-2\l} \left[\n^2 e^{4\b} + e^{-2\b} \left(
\nn e^{\sqrt{3} \bb} - \nnn e^{-\sqrt{3} \bb} \right)^2 \right .\nonumber\\
 &  &\left . - 2 \n e^{\b} \left(\nn e^{\sqrt{3} \bb} + \nnn
e^{-\sqrt{3} \bb} \right) \right] + \half \n\nn\nnn (1 + \n \nn \nnn ) \ ,
 \label{9} 
 \eeqa
for the types belonging to class A, and by
\beq
\RR = 2 a^2 e^{-2\l} \left( 3 - \frac{\nn\nnn}{a^2} \right) e^\beta \ ,
\label{10} 
\eeq
with
\beq
\beta = \frac{2}{3 a^2 - \nn \nnn} (\nn \nnn \b +  \sqrt{-3 a^2 \nn\nnn}\bb) 
\label{11} \ 
\eeq
for types in class B, with $ n_a^a = 0 $.
 
As said above, for types IV and VII$_{h \neq 0}$,
 the Lagrangian is not known, and therefore we shall not consider them here.

\section{\normalsize \bf N\"other symmetries}

According to a well established procedure \bib{demianski},\bib{deritis},
we look for
N\"other symmetries of the Lagrangian \num{8}. In other words we
look for a vector field $X$ on the configuration space, such that
the Lie derivative with respect to  the tangent 
lift $X^T$ of $X$ on the tangent
space vanishes
\beq
L_{X^T} \L = 0 \ .
\label{3.1}
\eeq
The conditions imposed by this equation generally puts restrictions on 
the class of allowed potentials for $\phi$.

But let us first observe that there are a number of trivial
cases.
\vspace{10pt}
\begin{itemize}
 \item 
 If $V = const.$, $\L$ does not depend on $\phi$, which is therefore
cyclic. Thus, for all classes and types, we always have the symmetry \beq
X = \D{\phi}  \ .
\label{3.2} 
\eeq
\item
 In Bianchi AI and BV, we have that $\RR$ does not depend on
$\b$, $\bb$, which are therefore cyclic. We have thus the symmetries \beq
X_1 = \D{\b}  \colon X_2 = \D{\bb} \ .
\label{3.3}
\eeq
Moreover, since the term $\bd^2 + \bbd^2$ is rotationally
invariant, we have the symmetry
\beq
X_3 = \b \D{\bb} - \bb \D{\b} + \bd \D{\bbd} - \bbd \D{\bd} \ .
\label{3.4}
\eeq
\item
 In Bianchi AII, $\RR$ depends on $\b$ only, so that we have the
symmetry \beq X = \D{\bb}\ .
\label{3.5}
\eeq
\end{itemize} In the following, using (\ref{3.1}), 
we investigate the possibility of
other symmetries, not so immediately evident.

First of all, we have to say that
the configuration space is made of four variables, namely $\l$,
$\b$, $\bb$ and $\phi$. Let us set
\beq
X = L \D{\l} + B_1 \D{\b} + B_2 \D{\bb} + F \D{\phi} \ ,
\label{3.6}
\eeq
where $L$, $B_1$, $B_2$ and $F$ are unknown functions of
$\l,\b,\bb,\phi$. The tangent lift will be
\beq
X^T =  X +  \frac{d L}{dt}  \D{{\dot \lambda}} +
\frac{d B_1}{dt} \D{\bd} +
\frac{d B_2}{dt} \D{\bbd} + \frac{d F}{dt} \D{\phidot} \ ,
 \label{3.7}
 \eeq
 where $ d/dt $ means Lie derivative along the dynamical vector
field; for example, it is
\beq
 \frac{d L}{dt}  = \frac{\d L}{\d \l}{\ld} + \frac{\d L}{\d
\b}{\bd} + \frac{\d L}{\d \bb}{\bbd} + \frac{\d L}{\d
\phi}{\phidot} \ .  \label{3.8}
\eeq
 
 The expression $ L_{X^T} \L $ is a homogeneous quadratic polynomial in
 $\ld,\bd,\bbd,\phidot$, plus a term of zeroth degree. Therefore, it
vanishes if and only if all the coefficients are zero
independently. This leads to a system of eleven partial differential
equations
\beq
3 L + 2 \frac{\d L}{\d \l} = 0 \colon
3 L + 2 \frac{\d B_1}{\d \b} = 0
\label{a}
\eeq

\beq 
3 L + 2 \frac{\d B_2}{\d \bb} = 0 \colon
 3 L + 2 \frac{\d F}{\d \phi} = 0
 \label{b}
 \eeq

\beq 
4 \frac{\d L}{\d \b} -  \frac{\d B_1}{\d \l} = 0 \colon
4 \frac{\d L}{\d \bb} -  \frac{\d B_2}{\d \l} = 0 
 \label{c}
 \eeq

\beq 
6 \frac{\d L}{\d \phi} -  \frac{\d F}{\d \l} = 0 \colon
 \frac{\d B_1}{\d \bb} +  \frac{\d B_2}{\d \b} = 0
 \label{d}
 \eeq

\beq 
3 \frac{\d B_1}{\d \phi} +2 \frac{\d F}{\d \b} = 0 \colon
3 \frac{\d B_2}{\d \phi} + 2  \frac{\d F}{\d \bb} = 0
 \label{e}
 \eeq

 \beq
 L \left(\RR + 6 V(\phi) + \n\nn\nnn(1 + \n\nn\nnn)\right) +
 B_1 \frac{\d \RR}{\d \b} + B_2 \frac{\d \RR}{\d \bb} + 2 F V' = 0
\ ,
 \label{f}
 \eeq
which are valid for both classes A and B, with relatively different expressions
for $\RR$.

We first obtain a general solution for equations (\ref{a}-\ref{e}), 
and then use
(\ref{f}) as a constraint to limit the class of solutions and  find
a condition on the potential.

From Eq.(\ref{a}a), we immediately get 
\beq
L = e^{-\ff \l} {\bar L}(\b,\bb,\phi) \ ,
\label{3.9}
\eeq
so that from (\ref{c}) 
\beq
B_{1,2} = -\frac{8}{3} \frac{\d {\bar L}}{\d \beta_{1,2}} e^{-\ff \l} +
{\bar B_{1,2}}(\b,\bb,\phi) \ ,
\label{3.10}
\eeq
and from (\ref{d}a)
\beq
F = -4 \frac{\d {\bar L}}{\d \phi} e^{-\ff \l}  + 
{\bar F}(\b,\bb,\phi) \ .
\label{3.11}
\eeq
Using (\ref{d}b), we get then
\beq
-\frac{8}{3} \frac{\d^2 {\bar L}}{\d \b \d \bb} e^{-\ff \l} +
\frac{\d {\bar B_1}}{\d \bb} = 
\frac{8}{3} \frac{\d^2 {\bar L}}{\d
\b \d \bb} e^{-\ff \l} - \frac{\d {\bar B_2}}{\d \b} \ .
\label{3.12}
\eeq

Since $ {\bar B_{1,2}}$
 do not depend on $\l$, it must be
\beq
 \frac{\d^2 {\bar L}}{\d \b \d \bb} = 0 \ ,
\label{3.13}
\eeq
so that
\beq
{\bar L} = {\bar L_1}(\b,\phi) + {\bar L_2}(\bb,\phi) \ ;
\label{3.14}
\eeq
moreover (\ref{d}b) gives
\beq
\frac{\d {\bar B_1}}{\d \bb} = - \frac{\d {\bar B_2}}{\d \b} \ .
\label{3.15}
\eeq
Substituting into (\ref{a}b), we get
\beq
3 e^{-3/2 \l}\left({\bar L_1} + {\bar L_2} -\frac{16}{9}
\frac{\d^2{\bar L_1}}{\d \b} \right) + \frac{\d {\bar B_1}}{\d \b} 
= 0 \ .
\label{3.16}
\eeq
Now, since only $\bar L_2$ and $\bar B_1$ depend on $\bb$, and $\bar B_1$
does not depend on $\l$, it must be
\beq
{\bar L_2} = 0 \colon \frac{\d {\bar B_1}}{\d \b} = 0 \ .
\label{3.17}
\eeq
Substituting into (\ref{b}a) we get analogously
\beq
{\bar L_1} = 0 \colon \frac{\d {\bar B_2}}{\d \bb} = 0 \ .
\label{3.18}
\eeq
Thus, we find that  $ L=0$ and $ B_1 = B_1(\bb,\phi)$, $B_2 = B_2(\b,\phi)$ \ .
 From (\ref{d}b) we obtain now
\beq
B_1 = f(\phi) \bb + g_1(\phi) \colon
B_2 = -f(\phi) \b + g_2(\phi) \ ,
\label{3.19}
\eeq
while from (\ref{b}b) and (7\ref{d}a) we get $ F = F(\b,\bb) $,
and from (\ref{e}a)
\beq
3 f'\bb + 3 g_1' = - 2 \frac{\d F}{\d \b} \ .
\label{3.20}
\eeq
The left hand side  does not depend on $\b$, thus
\beq
F = -\frac{3}{2} (g_1' - f'\bb)\b + {\tilde F}(\bb) \ ,
\label{3.21}
\eeq
and, analogously, from (\ref{e}b)
\beq
F = -\frac{3}{2} (g_2' - f'\b)\bb + {\hat F}(\b) \ .
\label{3.22}
\eeq

By comparison we get
\beq
f'=0 \colon g_1'' = g_2 '' = 0 \ ,
\label{3.23}
\eeq
\beq
-\frac{3}{2} g_1' \b - {\hat F{\b}} = - F_0 \colon
\frac{3}{2} g_2' \bb + {\tilde F{\bb}} =  F_0 \ ,
\label{3.24}
\eeq
where $F_0$ is an arbitrary constant.
The general solution is thus
\beqa
L=0 &\colon B_1 = c \bb + c_1 \phi + c_0 \\
 B_2 = -c \b + c_2 \phi + c'_0 &\colon 
 F = F_0 - \frac{3}{2} (c_1 \b + c_2 \bb) \ ,
 \label{3.25}
 \eeqa
with $c,c_1,c_2,c_0,c'_0,F_0$ arbitrary constants.

We have now to plug solutions \num{3.25}  into (\ref{f}) and check for
compatibility. Being $L=0$ we get
\beq
B_1 \frac{\d \RR}{\d \b} + B_2 \frac{\d \RR}{\d \bb}  = - 2 F V' \ .
\label{3.26}
\eeq

It is easy to see that the two sides must be zero independently, so
that we have two subcases: $V = const.$ and $ F = 0 $. In both of
them \num{3.26} is satisfied for types I and V. 
Examining the other types, we find that for Class A
Eq.\num{3.26}  is verified only for type II and $ B_1 = 0 $, while for Class B
Eq. \num{3.26} is verified for type VI$_{h \neq 0}$ iff
\def\aa{\frac{1}{a \sqrt{3}}}
\beq
c = 0 \colon c_2 = - \aa c_1 \colon
c'_0 =-\aa c_0 \ .
\label{3.27}
\eeq
The other types are excluded from the beginning, as said above.

\section{\normalsize \bf Results}
{\bf Case 1 -- $ V = costant $}

The situation for $ V = const. $ is summarized as follows

1a) Bianchi AI

The most general symmetry is written (on $Q$) as
\beq
X = (c \bb + c_1 \phi + c_0) \D{\b} + 
(- c \b + c_2 \phi+c'_0)\D{\bb} +
(F_0 - \frac{3}{2} (c_1 \b + c_2 \bb)) \D{\phi} \ ,
\label{4.1}
\eeq
with obvious lift to $TQ$. 
A basis of symmetries on $TQ$ is given by
\beq
X_1 = \D{\b} \colon X_2 = \D{\bb} \colon
X_3 = \D{\phi} 
\label{4.2}  
\eeq
\beq
X_4 = \phi \D{\b} - \frac{3}{2} \b \D{\phi} + \phidot \D{\bd} -
\frac{3}{2} \bd \D{\phidot}
\label{4.3}
\eeq
\beq
X_5 = \phi \D{\bb} - \frac{3}{2} \bb \D{\phi} + \phidot \D{\bbd} -
\frac{3}{2} \bbd \D{\phidot} \ .
\label{4.4}
\eeq
\beq
X_6 = \b\D{\bb} - \bb \D{b} + \bd \D{\bbd} - \bbd \D{\bd}  .
\label{4.4a}
\eeq
It turns out that only five of them are independent.
For instance, we can take the first five of them.
With
these symmetries, we associate the following constants of the
motion
\beq
K_1 = e^{3 \l} \bd \colon  K_2 = e^{3 \l} \bbd \colon
K_3 = e^{3 \l} \phidot
\label{4.5}
\eeq
\beq
K_4 = e^{3 \l} \left( \phi \bd - \frac{3}{2} \b \phidot \right)
\colon 
K_5 = e^{3 \l} \left( \phi \bbd - \frac{3}{2} \bb \phidot \right) 
\colon 
K_6 = e^{3 \lambda} (\b \bbd - \bd \bb) \ .
\label{4.6}
\eeq

These symmetries close on the following Lie algebra 
\beq
[X_i,X_j] = 0 \ \ , \ \ i,j = 1,2,3 \colon [X_1,X_4] =
-\frac{3}{2} X_3 \colon  [X_2,X_4] = 0
\label{4.7}
\eeq
\beq
[X_1,X_5]=0 \colon
[X_2,X_5] = -\frac{3}{2} X_3 \colon [X_3,X_4] = X_1 \colon [X_3,X_5] = X_2  
\label{4.9}
\eeq
\beq
 [X_4,X_5] = \frac{3}{2} X_6 \colon [X_1,X_6] = X_2 
\colon [X_2,X_6] = - X_1
 \label{4.10}
\eeq
\beq
[X_3,X_6] = 0 \colon [X_4,X_6] = X_5 \colon [X_5,X_6] = X_4 \ ,
\label{4.10a}
\eeq
and the $K_i$ close on the same algebra in terms of Poisson brackets.

This situation was examined in \bib{demianski} (with different variables). 
In that paper, it was shown how to use the symmetries in order to
obtain exact integration of the dynamics. We give an example below.

\noindent1b) Bianchi AII

The general form of $X$ (on $Q$) is
  \beq X = ( c_2 \phi + c_0)
\D{\bb} + \left(F_0 - \frac{3}{2} c_2 \bb\right) \D{\phi} \ ,
\label{4.11}
\eeq
from which we get the independent fields (on $TQ$)
\beq
X_1 = \D{\bb} \colon X_2 = \D{\phi} \colon
X_3 = \phi \D{\bb} - \frac{3}{2}\bb \D{\phi} + \phidot \D{\bbd} -
\frac{3}{2} \bbd\D{\phidot} \ ,
\label{4.12}
\eeq
with Lie algebra
\beq
[X_1, X_2] = 0 \colon  [X_1,X_3] = -\frac{3}{2} X_2 \colon
[X_2,X_3] = X_1 \ .
\label{4.13}
\eeq

The associated constants of the motion are a subset of the above
ones.

In spite of the existence of three symmetries, the exact
integration is not straitforward in this case, so that we shall not study
the situation here.

\noindent 1c) Bianchi BV

The situation is quite similar to Bianchi AI , with the same
symmetries and constants of the motion. The equations of course
are different. An example of exact integration is given below.

\noindent 1d) Bianchi BVI$_{h \neq 0}$

We have
\beqa
 X_1 &= &\D{\phi} \colon  X_2 = \D{\b} - \aa \D{\bb} \label{4.14}\\
 X_3 &= & \phi \D{\b} - \aa \phi \D{\bb} - \ff \left( \b -\ff \bb
\right) \D{\phi} + \nonumber\\ 
      & &\phidot \D{\bd} -\aa \phidot \D{\bbd} - \ff
\left( \bd -\ff \bbd \right) \D{\phidot} 
 \label{4.15} \eeqa
with algebra 
\beq
[X_1,X_2] = 0 \colon  [X_1,X_3] = X_2  \colon
[X_2,X_3] = \left( \frac{1}{3 a^2} -\ff \right) X_2
\label{4.16}
\eeq
and constants of the motion
\beq
K_1 = e^{3 \l} \phidot  \colon  K_2 = e^{3 \l} \left( \bd -\ff \bbd 
\right) 
\label{4.17}
\eeq
\beq
K_3 = e^{3 \l} \left( -3 \phi \b + \frac{1}{a^2} \phi \bbd + 3 \b 
\phidot - \frac{3}{a \sqrt{3}} \right) \ .
\label{4.18}
\eeq

\vspace{2. mm}

{\bf Case 2 --  $V \neq const.$ }

\noindent 2a) Bianchi AI

We obtain the following independent symmetries
\beq
X_1 = \D{\b} \colon X_2 = \D{\bb} \colon
X_3 = \b \D{\bb}  - \bb \D{\b} + \bd \D{\bbd} \bbd \D{\bd}
\label{4.19}
\eeq
with algebra
\beq
[X_1,X_2] = 0 \colon [X_1,X_3] = - X_2 \colon [X_2,X_3] = X_1
\label{4.20}
\eeq
and constants of the motion 
\beq
K_1 = e^{3 \l} \bd \colon K_2 = e^{3 \l} \bbd \colon
K_3 = e^{3 \l} ( \bb \bd - \b \bbd) \ .
\label{4.21}
\eeq
This case was examined already in \bib{demianski} and, although it was not
possible  to achieve exact integration, qualitative consideration
on the behaviour of the solutions were made possible.

 \indent 2b) Bianchi AII

Only one simmetry is left, namely
 \beq
X = \D{\phi}\;.
\label{4.22}
\eeq

\noindent 2c) Bianchi BV

Again the situation is essentially the same as for Bianchi AI.

\noindent 2d) Bianchi BVII$_0$

We have only one symmetry
\beq
X = \D{\b} -\ff \D{\bb}
\label{4.23}
\eeq
with the corresponding constant of the motion
\beq
K = e^{3 \l} \left( \bd - \aa \bbd \right).
\eeq

\section{\normalsize \bf Examples of exact integration}

In this Section, as an example, we obtain exact integration in the 
Bianchi AI and BV cases, when the potential $V(\phi)$ is zero.

As said earlier, the first order Einstein equation is a first integral 
of the second order equations derived from the Lagrangian \num{8}.
It turns out that it is always equivalent to 
\beq
E_\L \eqdef \frac{\d \L}{\d {\dot q}^i}{\dot q}^i - \L = 0 \ .
\label{5.1}
\eeq
It is in fact this equation that is left after reduction. Substituting
the constants of the motion found above we get
\beq
    6 {\dot \l}^2 - K^2 e^{-6 \l} - \alpha^2 e^{-2 \l} = 0 \ ,
    \label{5.2}
\eeq
where $ K^2 = 3(K_1^2 + K_2^2)/2 + K_3^2 $ and $ \alpha^2 = 6 a^2 e^ 
\frac{2}{3 a^2} $.

In Bianchi AI, i.e. when $ \alfa = 0$,  we obtain immediate integration
\beq
\l = \frac{1}{3} \log \left[\sqrt{\frac{3}{2}} K(t - t_0) \right] \ ,
\label{5.3}
\eeq
 which coincides with the one found in \bib{demianski}
and therefore, we avoid further comments on it.
In Bianchi BV, setting $ x = e^{2 \l}$ , we obtain the equation
\beq
\ff {\dot x}^2 -\frac{K^2}{x} - \alpha^2 = 0 \ ,
\label{5.4}
\eeq
which integrates to
\beq
\sqrt{\frac{2}{3}} t = \sqrt{x(1+x)} - \log (\sqrt{x} + \sqrt{1+x} ) \ ,
\label{5.5}
\eeq
where we have set $ K = 1$ (that is $ x(0) = 0 $), and $ \alpha = 1$ (that is, 
a rescaling of $x$ and $t$).

This relation cannot be inverted but can be used for a qualitative 
analysis. Indeed, it is easy to see that, for small $t$, and hence small $x$, 
we have that the mean scale factor $e^\lambda$ is proportional to $ t^{1/2}$, 
while for large $t$ it is proportional to $ t^{1/3}$. It is also possible to 
see that power law-inflation never occours.

As for the behaviour of $\phi$, it is immediately derivable from 
the expression of $K_3$. In the first case we get
\beq
        \phi = \sqrt{\frac{2}{3}} \frac{K_3}{K}\log (t-t_0) \ ,
\eeq
while in the second we have $ \phi \propto t^{-1/2} $ for small $t$ and
$ \phi \propto \log t $ for large $t$.

\section{\normalsize\bf Conclusions}

We  have found all N\"other symmetries for the Bianchi 
universes for which a Lagrangian function, of the type given in 
\bib{mccallum},
 is known. This allowed immediate exact  integration in simple cases.
 In the other cases we have seen that
 the number of symmetries is often
sufficiently high to permit a good reduction of the configuration 
space.  These situation are of great physical interest and a 
complete analysis  shall be done elsewhere.

It is important to  observe that the Lagrangian \num{8}, in the case 
when we want to add ordinary  matter (in form of dust decoupled from 
the scalar field), 
changes only by an additive constant. This is not irrelevant, because 
this constant changes the equation $ E_\L = 0 $, but of course the 
second order equations and the structure of the symmetries are exactly 
the same. This means that all the discussion above is valid for this 
case, except for the integration, which is now more complicate.

Other types of ordinary 
matter only give a contribution in the coefficient of $L$ in Eq.
\num{f}. Since $ L = 0$, we see that again the structure of symmetries is
 the same and the problems for integration may still 
arise from the energy condition.
\vspace{20pt}
\begin{center}
{\bf Acknowledgements:}
\end{center}
We want to thank our friend R. de Ritis for the many stimulating 
discussions about these topics.
\vspace{30pt}
\begin{center}
{\bf REFERENCES}
\end{center}
\begin{enumerate}
\item\label{kolb}
E.W. Kolb and M.S. Turner, {\it The early universe},
Addison-Wesley, New York (1990);\\ P.J.E. Peebles, {\it Principles of
physical cosmology}, Princeton Univ. Press, Princeton (1993).
\item\label{hoyle}
F. Hoyle and J.V. Narlikar, \prs {\bf A273} (1963) 1; \\
C. W. Misner, \apj {\bf 151} (1968) 431.
\item\label{wald}
R.M. Wald, \pr {\bf D28} (1983) 2118; \\
D.S. Goldwirt and T. Piran, \prep {\bf 214} (1992) 223.
\item\label{rothman}
T. Rothman and G.F.R. Ellis, \pl {\bf B180} (1986) 19.
\item\label{barrow}
J.D. Barrow, \pr {\bf D51} (1995) 3113.
\item\label{smoot}
G.F. Smoot and D. Scott, 
{\it Cosmic background radiation}, SISSA astro-ph/9603157.
\item\label{collins}
C.B. Collins and S.W. Hawking, \apj {\bf 181} (1972) 317.
\item\label{demianski}
M. Demianski, R. De Ritis, C. Rubano, P. Scudellaro and C. Stornaiolo, \pr
{\bf D46} (1992) 1391.
\item\label{Jantzen}
C. Uggla, R.J. Jantzen and K. Rosquist, \pr {\bf D51} (1995) 5522.
\item\label{lidsey}
J.E. Lidsey, {\it Symmetric vacuum 
scalar tensor cosmology}, SISSA gr-cq/9603052.
\item\label{deritis}
R. de Ritis, G. Marmo, G. Platania, C. Rubano, P. Scudellaro and C. Stornaiolo,
\pr {\bf D42} (1990) 1091;
\pl {\bf A149} (1990) 79;\\
S. Capozziello and R. de Ritis, \pl {\bf A177} (1993) 1; \ijmp {\bf
D2} (1993) 373; \cqg {\bf 11} (1994) 107; \ncim {\bf B109} (1994)
783; \ncim {\bf B109} (1994) 795;\\ S. Capozziello, R. de Ritis, C.
Rubano and P. Scudellaro, {\it N\"other symmetries in 
cosmology}, to appear in {\it La Rivista del Nuovo Cimento} (1996).
\item\label{ellis}
G.F.R. Ellis and M.A.H. McCallum, {\it Comm. Math. Phys.} {\bf 12} (1969) 108;\\
M.A.H. McCallum, in {\it Carg\`ese Lectures in Physics} {\bf Vol. 6}, ed. 
E. Schatzmann, Gordon \& Breanch, N.Y. (1973).
\item\label{bianchi} L. Bianchi, {\it Mem. Soc. It. della Sc. (dei XL)}
{\bf 11} (1897) 267; {\it Lezioni sulla teoria dei gruppi continui
finiti di trasformazioni}, Spoerri, Pisa (1918). 
\item\label{marmo1}
J.F. Cari\~nena, L.A. Ibort, G. Marmo and A. Perelomov, J. Phys. A: Math. Gen.
{\bf 27} (1994) 7425.
\item\label{marmo2}
J. Grabowski, G. Marmo and A. Perelomov, Modern Phys Lett. {\bf A8} (1993) 
1719.
\item\label{mccallum}
M.A.H. McCallum, in {\it General relativity: an Einstein centenary survey}, eds.
S.W. Hawking and W. Israel, Cambridge Univ. Press, Cambridge (1979).
\item\label{ryan}
M.P. Ryan Jr., \jmp {\bf 15} (1974) 812;\\
M.P. Ryan Jr. and L. Shepley, {\it Homogeneous relativistic cosmologies}, 
Princ. Univ. Press, Princeton (1975). 

\end{enumerate}

\end{document}